\documentclass[12pt]{iopart}

\usepackage[squaren]{SIunits}
\usepackage{stmaryrd}
\usepackage{amsthm}
\usepackage{amssymb}
\usepackage{mathrsfs}
\usepackage{tensor}

\usepackage{multirow}
\usepackage{color}
\usepackage{setspace}

\usepackage{graphicx}
\usepackage{subfig}

\usepackage{url}

\usepackage{hyperref}

\begin{document}

\title[Equatorial orbits and imaging of hairy cubic Galileon black holes]{Equatorial orbits and imaging of hairy cubic Galileon black holes}
\author{K.~Van Aelst$^{1,2}$, E.~Gourgoulhon$^{2}$, F.~H.~Vincent$^{3}$}
\vspace{10pt}
\begin{indented}
\item[$^{1}$] Max Planck Institute for Gravitational Physics (Albert Einstein Institute), Am~M\"{u}hlenberg 1, 14476 Potsdam, Germany
\item[$^{2}$] Laboratoire Univers et Th\'eories, Observatoire de Paris, Universit\'e PSL, CNRS, Universit\'e de Paris, 5 place Jules Janssen, 92190 Meudon, France
\item[$^{3}$] LESIA, Observatoire de Paris, Universit\'e PSL, CNRS, Sorbonne Universit\'e, Universit\'e de Paris, 5 place Jules Janssen, 92190 Meudon, France
\end{indented}
\ead{karim.vanaelst@aei.mpg.de, eric.gourgoulhon@obspm.fr, frederic.vincent@obspm.fr}

\begin{abstract}
Null and timelike equatorial orbits are investigated in a family of hairy black holes in the cubic Galileon theory.
These include rotating generalizations of static black hole metrics supporting a time-dependent scalar field.
Depending on the coupling and rotation, the properties of the geodesics expectedly deviate from general relativity.
In particular, it is found that stable circular geodesics only exist below a critical coupling, which is related to the existence of an outermost stable circular orbit.
Focusing on the strong-field region, images of an accretion disk are also produced to highlight tendencies that would constrain the model given further accurate observations of supermassive black holes.
\end{abstract}


\section{Introduction}

Trajectories of stars and images of accretion disks orbiting black holes provide some of the main observables to test strong-field gravity~\cite{Bambi_BH_test_strong_grav_book,Cardoso_review_test_compact}.
Such observational data are collected by complex instruments like the interferometer GRAVITY~\cite{GRAVITY}\nocite{GRAVITY_redshift_S2, GRAVITY_motion_ISCO} and the Event Horizon Telescope (EHT)~\cite{EHT}\nocite{EHT_Shape_SgrA, EHT_Shadow_M87}, which mainly focus on the supermassive black holes~\emph{Sgr~A*} and~\emph{M87*}~\cite{GRAVITY_redshift_S2, GRAVITY_motion_ISCO,EHT_Shape_SgrA, EHT_Shadow_M87}.
The theoretical predictions for these observables have been worked out within more or less exotic frameworks: Kerr black holes (e.g. \cite{Vincent_model_M87_2020}), rotating black holes dressed with a complex scalar hair~\cite{image_Kerr_scalar_hair, Collodel21_circular_orbs_thin_disk_KerrScalar}, boson stars~\cite{Grould17_geods_boson, image_boson, Olivares20_accret_boson, Vincent_model_M87_2020},
alternative black holes~\cite{Vetsov18_shadow_vect_Gal, Wang18_shadow_non_Kerr, Tsukamoto18_shadow_regular, Wang19_shadow_Kerr_like_MOG, Bambi19_shadow_bh_mimicker, Moffat20_shadow_MOG, Khodadi20_shadow_minim_and_conform, geods_shadow_GB, Wei20_shadow_4DEGB_bh, Guo20_ISCO_shadow_4DEGB_bh}, wormholes~\cite{image_regular_BH, Vincent_model_M87_2020}, naked singularities~\cite{Shaikh19_image_naked_singu}, binary systems~\cite{Cunha18_shadow_double_Sch, Cunha18_shadow_exact_bbh}.
Eventually, such analyses help constraining the nature of the observed objects~\cite{Cardoso_review_test_compact}, but also the theory of gravity within which they are modeled~\cite{Psaltis20_post_newt_tests_M87, Volkel20_EHT_strong_GR, EHT21_Constraints_BHcharges_M87}.

It must be noted though that unequivocally excluding a given model is a complex endeavour in most cases.
Current instrumental limitations and extremely large parameter spaces (describing e.g. the emission flow or sources of noise) require to rely on simplifying prescriptions~\cite{THEMIS20_Parameter_EHT} to realize and interpret observations, such as the EHT reconstructed image.
This leaves too much uncertainty to draw definitive conclusions today, and extensive studies are needed to explore significant parts of the parameter spaces and guess which observations could be decisive in discriminating some given models.
But the understanding of astrophysical black holes gradually progresses by improving current instruments and analysis tools, and developing ideas for future enhanced observations.

In this context, the present paper focuses on the characteristic features and preliminary constraints arising from the strong field observables of a family of black hole spacetimes within the cubic Galileon theory.
The (``covariant generalized'') Galileons are scalar-tensor theories which coincide with Horndeski theories in four dimensions, meaning that they are the most general scalar-tensor theories leading to second-order field equations~\cite{Horndeski_original, Galileon_original, covariant_Galileon, generalized_Galileons, G_inflation, Deffayet_review_Galileon}.
Galileons thus provide a relevant framework to search for observable deviations from general relativity (GR), as most signatures of alternative theories are expected to be well described by scalar-tensor theories at least in some effective range.
In particular, the cubic Galileon emerges from effective formulations of higher dimensional theories, either in the decoupling limit of braneworld models such as the popular 5-dimensional Dvali-Gabadadze-Porrati (DGP) model~\cite{DGP_original, DGP_effective_action_2, Galileon_original, de_Rham_review_Galileons, de_Rham_review_massive}, or from Kaluza-Klein compactification of Lovelock theory~\cite{Galileons_Lovelock, Christos_Lovelock_Horndeski}.
Explicitly, its vacuum action writes
\begin{eqnarray}
\label{eq_action}
S\left[ g,\phi \right]
&= \int \left[
            \zeta (R - 2 \Lambda)
            - \eta (\partial \phi)^{2}
            + \gamma (\partial \phi)^{2} \Box\phi
        \right]
    \sqrt{\vert \det g \vert} d^{4}x,
\end{eqnarray}
where~$(\partial \phi)^{2} \equiv \nabla_{\mu} \phi \nabla^{\mu} \phi$ and~$\zeta$,~$\eta$ and~$\gamma$ are coupling constants.

It is the simplest of Galileons with higher order derivatives, and it is compatible with the observed speed of gravitational waves \cite{GW_speed, DE_after_GW_Vernizzi, DE_after_GW_ahead, DE_after_GW_revisited}.
As a well-motivated, consistent theory, the cubic Galileon has been investigated in various contexts, from laboratory tests \cite{Galileon_labo} to cosmology \cite{Galileon_linear_cosmo_perturb, Galileon_cosmo_viability, cubic_Galileon_structure_formation, time_dep_cubic_cosmo, cosmo_self_tun}.
Besides current observations of supermassive black holes, further interest in the characteristics of cubic Galileon black holes comes from the fact that, together with most shift-symmetric
\footnote{
Shifting the scalar field by a constant ($\phi\rightarrow\phi+constant$) preserves action~(\ref{eq_action}).
} 
Horndeski theories, the cubic Galileon is subject to a ``no-scalar-hair'' theorem: in the asymptotically flat framewok, the only static, spherically symmetric black hole metric and scalar field are the Schwarzschild metric together with a trivial scalar field~\cite{no_hair_Galileon, hairy_BH_GB_Sotiriou_1, hairy_BH_GB_Sotiriou_2, Babichev_BH_stars_Horndeski} (see also~\cite{slowly_rotating_no_hair} for an extension to slow rotation and~\cite{Christos_no_hair_stars_Horndeski} for stars).
Consequently, modified gravity effects can only occur in systems breaking one of these hypotheses.
Indeed, one does obtain non-GR metrics coupled to non-trivial scalar hair when enforcing all hypotheses except the stationarity of the scalar field.
Such minimal violation of the hypotheses is possible when the scalar field features a linear time-dependence~\cite{Christos_BH_time_dep_Gal}
\begin{eqnarray}
\label{eq_scalar_ansatz}
\phi = qt + \Psi,
\end{eqnarray}
where~$q$ is a non-zero constant and~$\Psi$ is time-independent.

The shift-symmetry is what makes a linear time-dependence compatible with a static and spherically symmetric metric.
Such configuration was considered in various contexts such as cosmology, and the linear time-dependence was physically interpreted as a first order approximation to a slowly evolving scalar field~\cite{Galileon_accretion, constraints_time_variation_G_qt, time_dep_cubic_cosmo, cosmo_self_tun}.
Besides, using ansatz~(\ref{eq_scalar_ansatz}) (with~$\Psi$ depending not only on the radial but also the angular coordinate), previous numerical work~\cite{rot_cubic_BH} produced rotating generalizations of hairy static and spherically symmetric solutions derived in the cubic Galileon theory~\cite{cubic_BH}.
At the level of the metric, these rotating black holes significantly deviate from Kerr spacetime, implying possibly observable modified gravity effects in black hole environments.

Such hairy configurations were constructed as circular spacetimes, meaning that they were assumed to admit a quasi-isotropic coordinate system with respect to which the line element writes
\begin{eqnarray}
\label{eq_circular_spacetime}
ds^{2} = - N^2 dt^2 + A^2 \left( dr^2 + r^2 d\theta^2 \right) + B^2 r^2 \sin^2 \theta \left( d\varphi - \omega dt \right)^2.
\end{eqnarray}

Circular spacetimes represent a large subclass of stationary and axisymmetric spacetimes, and their quasi-isotropic coordinates are well suited to study geodesics.
Yet compatibility of a circular metric~(\ref{eq_circular_spacetime}) with a linear time dependence~(\ref{eq_scalar_ansatz}) is exact only in the non-rotating case, while errors arise as rotation increases and could become significant at high rotation.
This is why these spacetimes are only considered at low and moderate rotation such that the errors on the solutions are negligible~\cite{rot_cubic_BH}.
Despite such restriction, these configurations allow to observe non-perturbative effects of rotation (examples of which are still not so abundant in modified gravity, although relevant for astrophysical black holes which are expected to rotate).

Besides, the metrics were constructed to be asymptotically flat, as it is a natural hypothesis of the no-scalar-hair theorem.
Yet this is realized by taking~$\eta = \Lambda = 0$ in the cubic Galileon model~(\ref{eq_action}), leaving the scalar field ruled by the non-standard ``DGP term''~$(\partial \phi)^{2} \Box\phi$.
This induces a non-standard asymptotic behaviour of the metric which yields a vanishing Komar mass at infinity.
This theoretical fact is unusual, but astrophysically, it does not forbid the hole to feature an effective mass on finite distances.
More precisely, we will see that there do exist stable orbits, yet only up to an OSCO (outermost stable circular orbit).
OSCO's are not so unusual even in GR in presence of a positive cosmological constant, and they can emerge on short enough scales to be astrophysically relevant (for instance, using the Schwarzschild-de Sitter metric outside of a galaxy, the OSCO is of the order of the inter-galactic distance~\cite{Visser_ISCO_OSCO_Lambda}).
In the cubic Galileon case, studying the geodesics of the asymptotically flat hairy black holes will show that the OSCO actually emerges on an even shorter scale when the remaining parameters of the model are chosen so as to generate non-negligible metric deviations in the strong-field region.
The asymptotically flat model itself is then strongly constrained by requiring stability of all the orbits that would be dominantly ruled by a central supermassive black hole.

Note however that non-zero couplings~$\eta$ and~$\Lambda$ lead to Schwarzschild-de Sitter asymptotics in the static case~\cite{cubic_BH}, which generically restore a large OSCO compatible with observations.
Despite different asymptotic properties, models with non-zero~$\eta$ and~$\Lambda$, and asymptotically flat solutions, might present analogous dependencies on the DGP coupling and on rotation, and share common observable characteristics in the strong field region.
This is why images of an accretion disk orbiting the asymptotically flat black holes are also presented below, while future work will construct asymptotically de Sitter solutions to assess such similarities (and more generally to identify degeneracies of the observables within the cubic Galileon model).

Besides, if some strong field characteristics of the asymptotically flat configurations turned out particularly interesting without being preserved for any non-zero~$\eta$ and~$\Lambda$, one might deal with the unusual large distance properties of the asymptotically flat model by relying on convenient mechanisms or fields, such as a second, ``screened'' scalar field~$\chi$.
Screened scalar fields are commonly invoked in cosmology, in different realizations of massive gravity and hence in Galileon theories~\cite{de_Rham_review_Galileons} to recover the successful predictions of GR on short scales, e.g. solar system scales, while providing new relevant cosmological phenomenology in regard of dark energy.
In standard screening mechanisms (Chameleon~\cite{2004_original_Chameleon, 2007_Mota_review_Chameleon}, Symmetron~\cite{2008_Olive_Symmetron, 2010_Hinterbichler_Symmetron}, Vainshtein~\cite{original_kMouflage, Hinterbichler_review_massive, Babichev_review_Vainshtein, de_Rham_review_massive}), the mass of~$\chi$ or its coupling to matter effectively decrease in regions of high matter density.
Such processes are realized through non-standard kinetic terms and/or appropriate couplings to standard model matter, such as Dirac spinors, Higgs scalars or other fundamental fields~\cite{2008_Olive_Symmetron}.
Although the idea has not been considered elsewhere, these mechanisms might be adapted to the present asymptotically flat case, e.g. by introducing analogous interactions that suitably couple~$\chi$ to the black hole hair~$\phi$ (rather than to standard matter): $\chi$ would be screened where~$\phi$ adopts its strong field profile, i.e. the immediate vicinity of the black hole, while it would modify the geometry on larger distances, and in particular the location of the OSCO.
Thus, the asymptotically flat case may provide an effective description of the strong field region of more complete models that feature more standard weak field properties thanks to a standard kinetic term for~$\phi$, a cosmological constant~$\Lambda$, or alternative fields or mechanisms.

The plan of the article is as follows.
The properties of the equatorial timelike circular geodesics and photon rings (location, stability, deviation from Kerr spacetime) are studied in the static, spherically symmetric case in section~\ref{section_geod_stat}, and the rotating case in section~\ref{section_geod_rot}.
This leads to to compute the images of an accretion disk orbiting the static, spherically symmetric black holes in section~\ref{section_im_rot}, and the rotating black holes in section~\ref{section_im_rot}.
Notations and general results on equatorial geodesics in quasi-isotropic coordinates are summarized in appendix~\ref{appdx_geod_QI}.

\section{Orbits around cubic Galileon black holes}
\label{section_geod_num}

\subsection{Static and spherically symmetric case}
\label{section_geod_stat}

\begin{figure}
    \begin{center}
        \subfloat[Radial profile of function~$D$ whose positivity allows (possibly superluminal) circular geodesics.]
        {
            \includegraphics[width=0.48\textwidth]{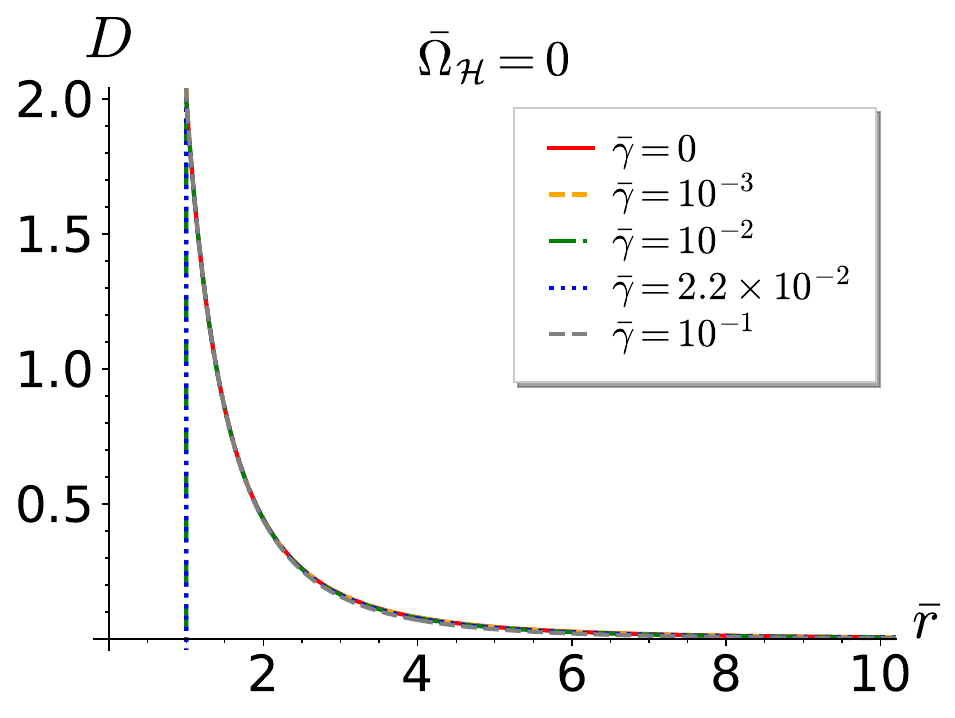}
            \label{fig_D_stat}
        }
        \subfloat[Velocities and photon rings.]
        {
            \includegraphics[width=0.49\textwidth]{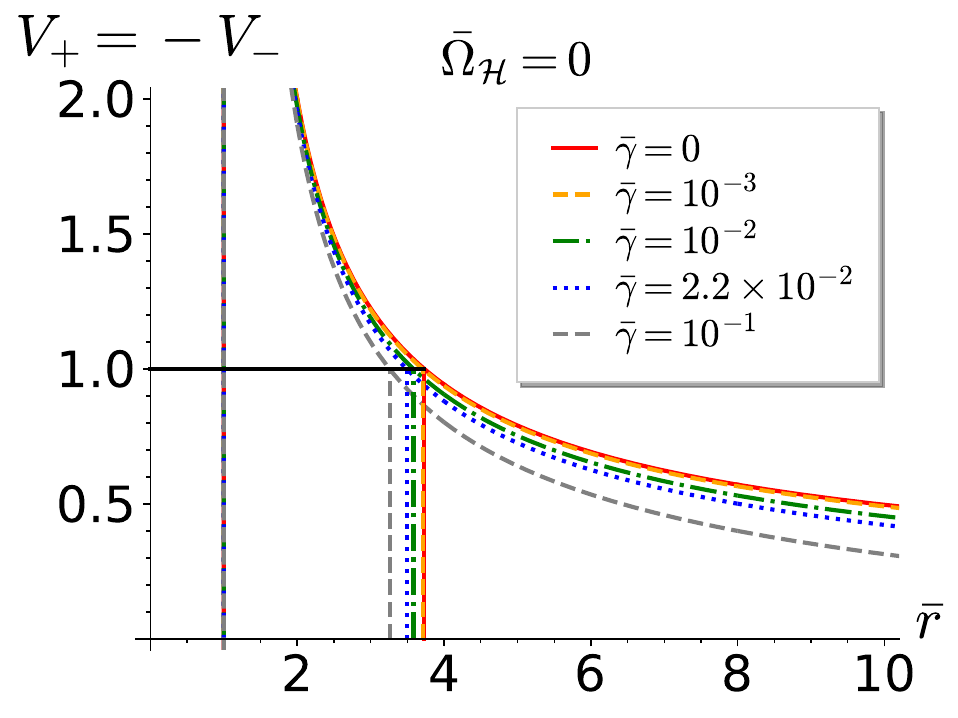}
            \label{fig_V_stat}
        }
    \end{center}
\caption{Radial profiles of~$D$ and the resulting velocities of circular geodesics in the static, spherically symmetric case~(i.e. vanishing dimensionless angular velocity of the event horizon~$\bar{\Omega}_{\mathcal{H}} = r_{\mathcal{H}} \Omega_{\mathcal{H}}$) for various couplings ($\bar{\gamma}=0$ corresponding to Schwarzschild spacetime). The lapse~$N$ and its derivative are positive everywhere (see figure 2(a) of~\cite{rot_cubic_BH}), so that~$D > 0$ implies that the denominator in~(\ref{eq_V_pm}) is positive. Therefore,~$V_{+} > 0$ (prograde orbits) and~$V_{-} = - V_{+} < 0$ (retrograde orbits).}
\label{fig_stat_sphe}
\end{figure}

To study the geodesics of the cubic Galileon static and spherically symmetric black holes obtained in~\cite{rot_cubic_BH,cubic_BH}, the procedure is to first characterize the circular geodesics.
As detailed in appendix~\ref{appdx_geod_QI}, regions of positive discriminant~$D$~(\ref{eq_D}) are first checked on figure~\ref{fig_D_stat} for different values of~$\bar{\gamma} = q^{3} r_{\mathcal{H}} \gamma/\zeta$, where $r_{\mathcal{H}}$ is the radial coordinate of the event horizon
\footnote{
Recall that~$\bar{\gamma}$ is the only remaining dimensionless coupling parametrizing deviations because~$\eta$ and $\Lambda$ are set to $0$; the values of~$\bar{\gamma}$ considered in figure~\ref{fig_D_stat} are picked in the range leading to non-negligible metric deviations from GR~\cite{rot_cubic_BH}.
}
.
Function~$D$ appears positive everywhere down to the horizon, where it diverges because of division by the lapse~$N$ which cancels at the horizon.
Therefore, circular geodesics a priori exist everywhere for all couplings, but they necessarily become superluminal near the horizon according to~(\ref{eq_V_pm}).

This is what figure~\ref{fig_V_stat} confirms: for each coupling~$\bar{\gamma}$, velocity diverges at the horizon so that there exists a photon ring (marked with a vertical line from~$0$ to~$1$), only beyond which timelike circular geodesics exist.
Although~$D$ does not vary much with coupling on figure~\ref{fig_D_stat}, the velocities more strongly depend on~$\bar{\gamma}$ because function~$B$ in the denominator of~(\ref{eq_V_pm}) does vary (see figure 2(b) of~\cite{rot_cubic_BH}, knowing that~$B = A$ everywhere in spherical symmetry).
More precisely, at fixed radius, the velocity of the circular geodesic decreases with increasing coupling.
As a consequence, the photon ring gets closer to the horizon as~$\bar{\gamma}$ increases.

These results are related to the following facts mentioned in the introduction.
The metric functions~$N$ and~$A = B$ converge faster to Minkowski at infinity as~$\bar{\gamma}$ increases~\cite{rot_cubic_BH}.
Therefore, at fixed radius away from the strong-field region, gravitation gets naively weaker as~$\bar{\gamma}$ increases, so that the velocity of the circular geodesic must be smaller.
In addition, for any~$\bar{\gamma} \neq 0$, convergence to Minkowski spacetime is always much faster than that of Schwarzschild spacetime:~$N$ and~$A = B$ converge to~$1$ as~$1/r^{4}$ rather than~$1/r$, yielding a vanishing Komar mass at infinity \cite{rot_cubic_BH}.
As a result, velocities given by~(\ref{eq_V_pm}) converge to zero like~$r^{-\alpha/2}$ with~$\alpha = 1$ in Schwarzschild spacetime and~$\alpha = 4$ in Galileon spacetimes.

\begin{figure}
    \begin{center}
        \subfloat[Lorentz factor.]
        {
            \includegraphics[width=0.5\textwidth]{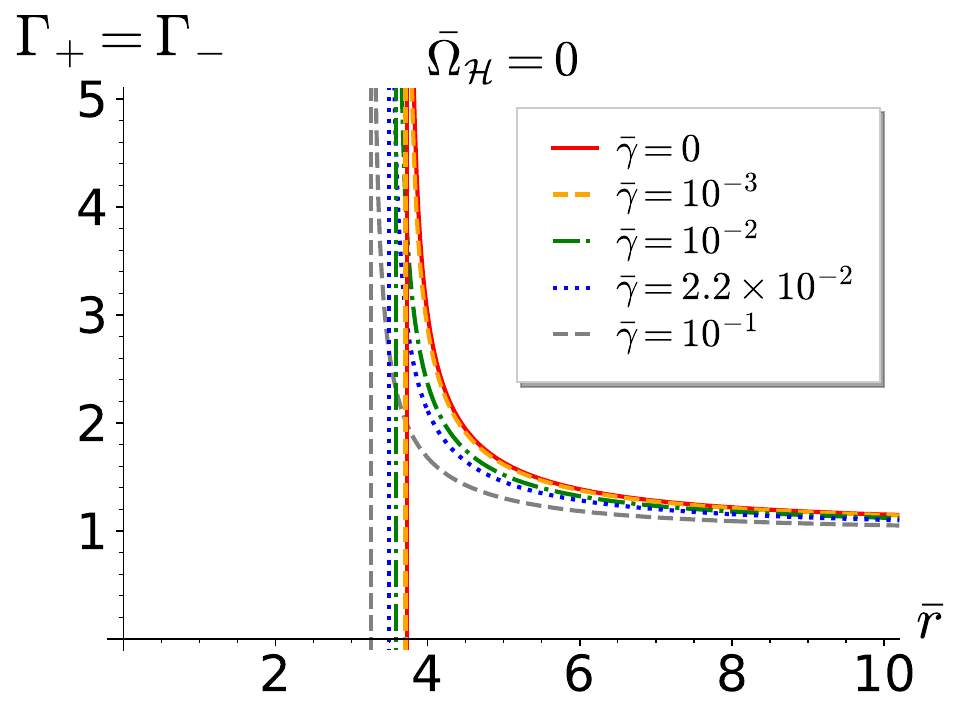}
            \label{fig_Gamma_stat}
        } \\
        \subfloat[Killing angular momentum.]
        {
            \includegraphics[width=0.49\textwidth]{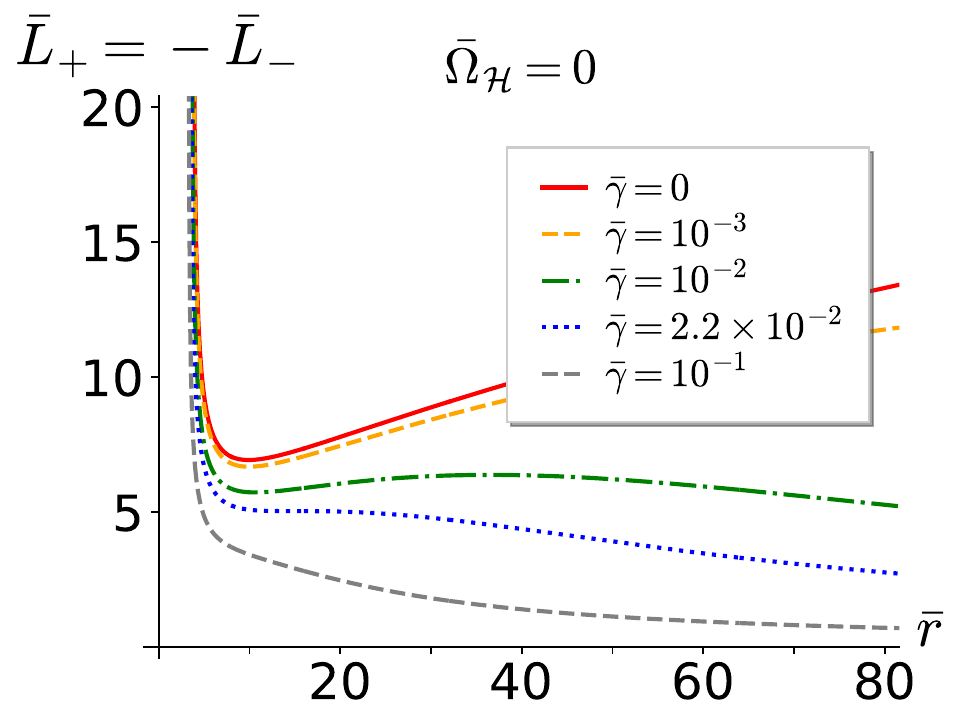}
            \label{fig_L_stat}
        }
        \subfloat[Killing energy.]
        {
            \includegraphics[width=0.48\textwidth]{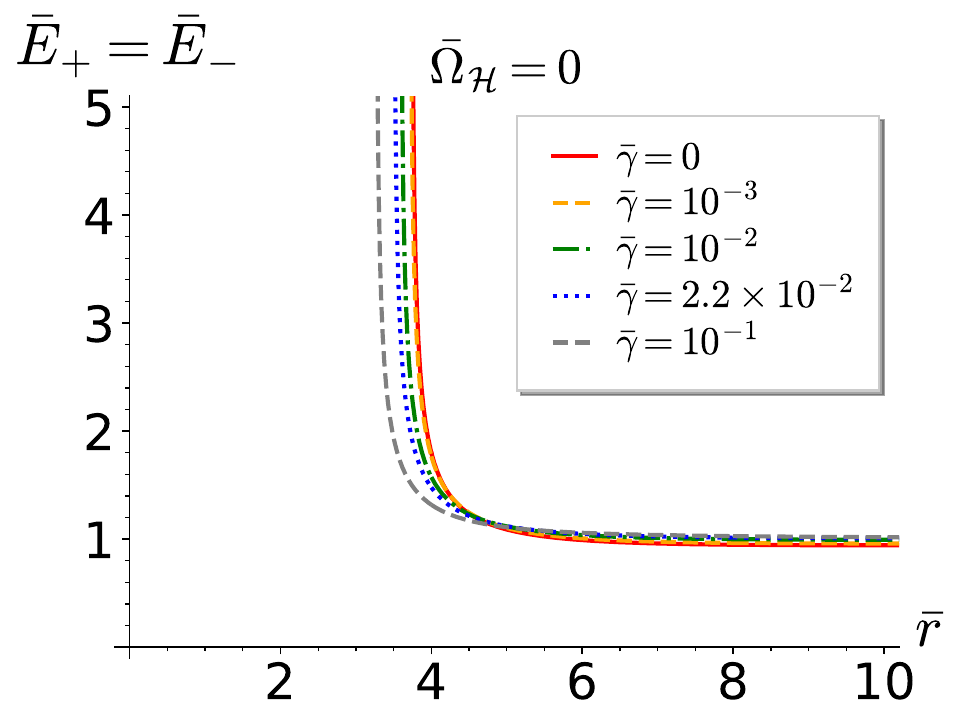}
            \label{fig_E_stat}
        }
    \end{center}
\caption{Radial profiles of kinematic quantities measured by the ZAMO for the timelike circular geodesics in the static, spherically symmetric case for various couplings. They all diverge at the photon ring (yet asymptotes are only plotted for the Lorentz factor).}
\label{fig_kinema_stat_sphe}
\end{figure}

Such asymptotic behaviours are highlighted in figures~\ref{fig_kinema_stat_sphe}.
In all cases, the Lorentz factor displayed in figure~\ref{fig_Gamma_stat} logically converges to~$1$.
However, according to~(\ref{eq_L_circu_V}), the Killing angular momentum per unit mass~$\bar{L}$ displayed in figure~\ref{fig_L_stat} behaves like~$rV \simeq r^{1-\alpha/2}$, hence the divergence in Schwarzschild spacetime and convergence to zero for any~$\bar{\gamma} \neq 0$
\footnote{
The numerical solutions contain information at infinity confirming this fact even for small couplings like~$\bar{\gamma} = 10^{-3}$ whose convergence to zero becomes apparent very far from the horizon.
}
.
Finally,~$\bar{E} = \Gamma N$ converges to~$1$ in all cases on figure~\ref{fig_E_stat}, which will hold true in the rotating case since function~$\omega$ will converge to~$0$ like~$1/r^{3}$ regardless of whether~$\bar{\gamma}$ is zero or not.
At the photon ring, all the kinematic quantities displayed in figure~\ref{fig_kinema_stat_sphe} naturally diverge.

\begin{figure}
    \begin{center}
    \includegraphics[width=0.5\textwidth]{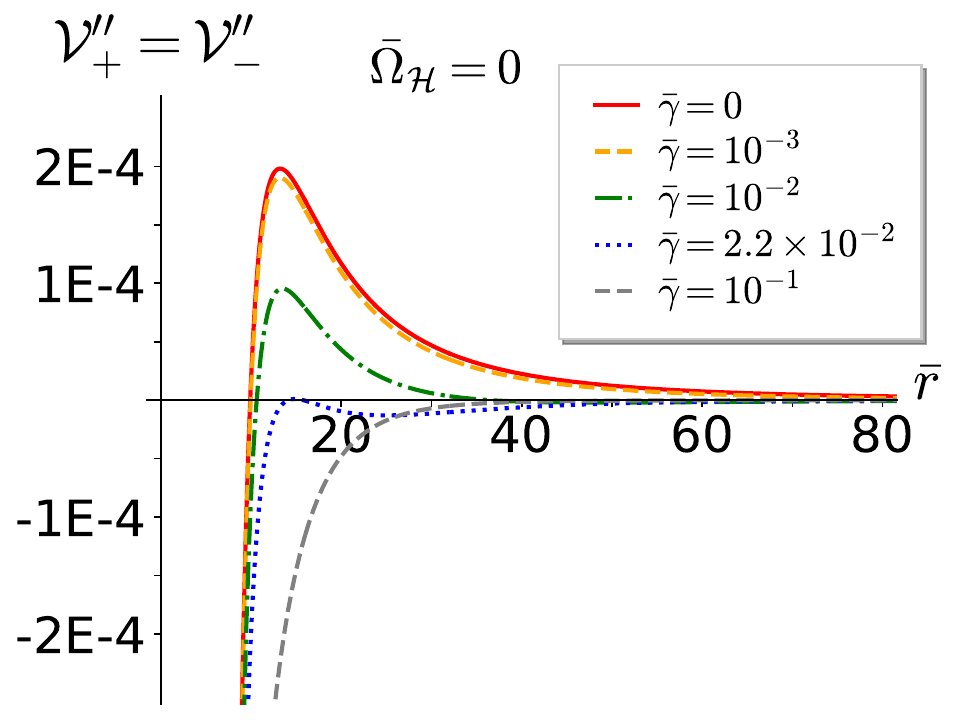}
    \end{center}
    \caption{Radial profile of~$\mathcal{V}_{\pm}''$; positivity determines stability of the geodesics.}
\label{fig_drrPot_stat}
\end{figure}

Figure~\ref{fig_drrPot_stat} assesses the stability of circular orbits for various couplings, based on the functions~$\mathcal{V}_{\pm}''$ given by~(\ref{eq_sign_stab}).
As explained in appendix~\ref{appdx_geod_QI},
their sign at a given radial coordinate~$r_{0}$ is the same as~$\mathcal{V}''\left(r_{0},1,\bar{E}_{\pm}(r_{0}),\bar{L}_{\pm}(r_{0})\right)$ respectively.
It appears that for any non-zero~$\bar{\gamma}$, both an innermost and an outermost stable circular orbit (ISCO and OSCO) exist.
They respectively correspond to the smallest and greatest~$r_{0}$ such that~$\mathcal{V}_{\pm}''(r_{0}) = 0$.
Since~$\mathcal{V}_{\pm}''$ globally decreases as~$\bar{\gamma}$ increases, the ISCO radius increases while the OSCO decreases from infinity (where it is formally located in the Schwarzschild case~$\bar{\gamma} = 0$).
In particular, the photon ring is always unstable because its location decreases when~$\bar{\gamma}$ increases (figure~\ref{fig_V_stat}), so that it remains below the ISCO as in Schwarzschild spacetime.

The ISCO and OSCO eventually merge for a critical coupling~$\bar{\gamma}^{c}~\simeq~2.2~\times 10^{-2}$, beyond which no stable circular orbit exists anywhere.
Therefore, the mere existence of stars orbiting black holes such as~\emph{Sgr~A*} in a seemingly stable way requires~$\bar{\gamma} \ll \bar{\gamma}^{c}$ for the present static Galileon black hole to be viable.
The existence of a particularly close OSCO further constrains the model since e.g. the well-known star~S2 lies beyond~$2500 M$, where~$M \simeq 4 \times 10^{6} M_{\odot}$ is the observed mass of~\emph{Sgr A*}.
Although its orbit is non-circular, it indicates that a stable circular orbit exists between its apsides, and hence the OSCO must lie beyond.
For instance, this further requires~$\bar{\gamma} \ll 10^{-3}$.

Note though that a perturbation~$\delta$ away from a circular orbit at some radius~$r_{0} > r_{OSCO}$ obeys equation
\begin{eqnarray}
\ddot{\delta} + \mathcal{V}''\left(r_{0},1,\bar{E}_{\pm}(r_{0}),\bar{L}_{\pm}(r_{0})\right)\delta = 0,
\end{eqnarray}
so that the instability timescale is
\begin{eqnarray}
\tau(r_{0}) = 1/\sqrt{-\mathcal{V}''\left(r_{0},1,\bar{E}_{\pm}(r_{0}),\bar{L}_{\pm}(r_{0})\right)}.
\end{eqnarray}
Since~$\mathcal{V}''$ asymptotically goes to zero as~$r^{-6}$, $\tau$ diverges and rapidly becomes larger than the age of the universe (e.g. around~$30 r_{OSCO}$ for $r_{\mathcal{H}}$ of the order of the Schwarzschild radius of~\emph{Sgr A*} and for~$\bar{\gamma} \sim 10^{-2}$).
Although this does not render the model viable
\footnote{
This would not allow to observe today distant bounded eccentric orbits like S2, but only quasi-circular orbits very slowly drifting away which would have never experienced any strong radial perturbation.
}, 
such instability is weaker than in the standard cases featuring an OSCO (in Schwarzschild-de Sitter spacetime, $\tau$ converges to a finite value, fixed by the cosmological constant).
This goes along with the fact that this close OSCO is a singular artefact of the marginal combination~$\eta = \Lambda = 0$, and corroborates the idea that a much larger stability region would be restored with any additional Lagrangian terms or mechanism mentioned in the introduction.
With this in mind, the next section focuses on the effects of rotation on the orbits.

\subsection{Rotating case}
\label{section_geod_rot}

\begin{figure}
    \begin{center}
        \subfloat[Velocities ($V_{+} > 0,\ V_{-} < 0$) and photon rings.]
        {
            \includegraphics[width=0.49\textwidth]{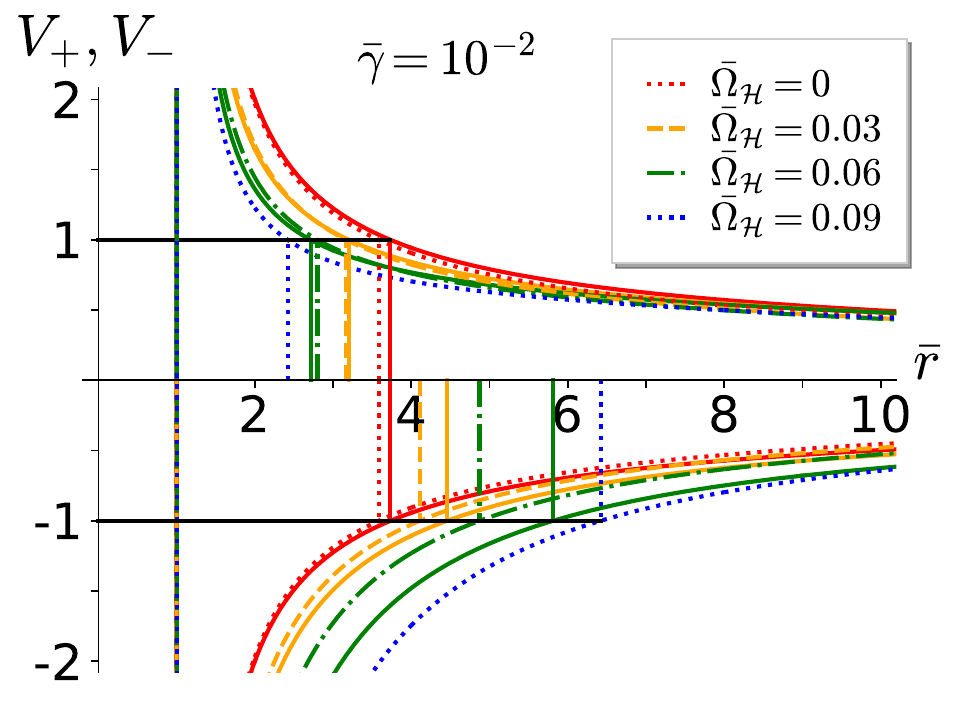}
            \label{fig_V_rot}
        }
        \subfloat[Angular momentum ($\bar{L}_{+} > 0,\ \bar{L}_{-} < 0$).]
        {
            \includegraphics[width=0.49\textwidth]{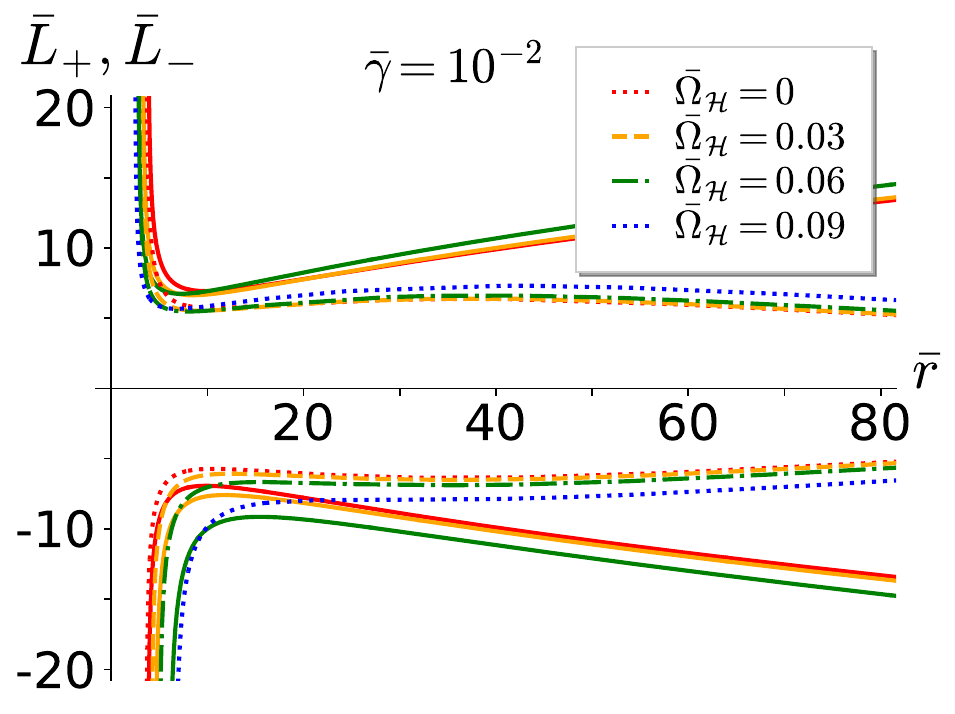}
            \label{fig_L_rot}
        } \\
        \subfloat[Energy of prograde geodesics.]
        {
            \includegraphics[width=0.49\textwidth]{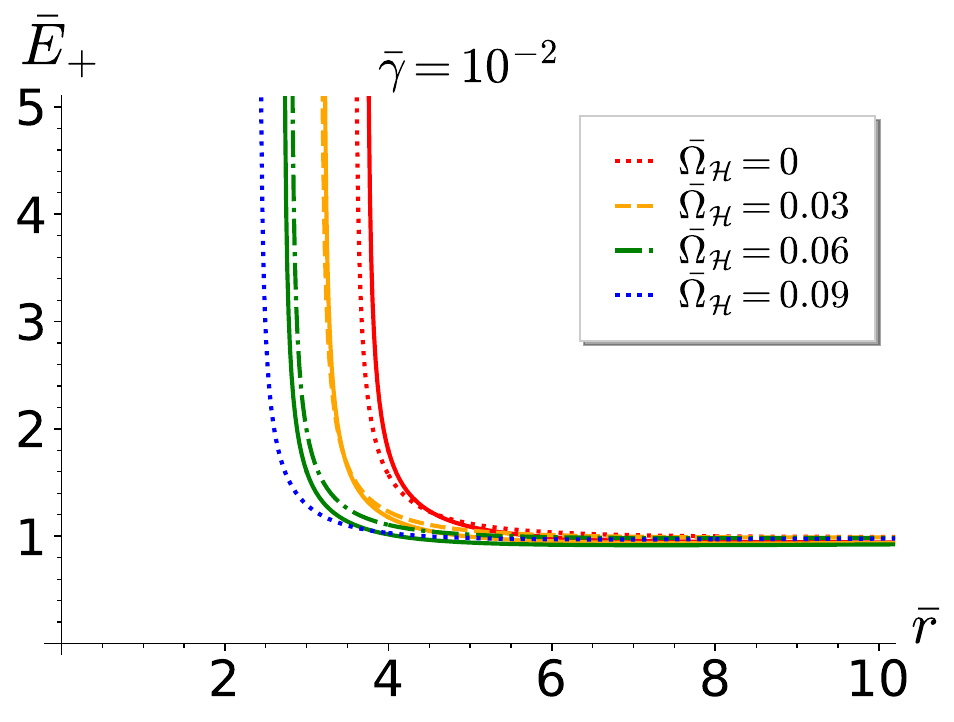}
            \label{fig_Ep_rot}
        }
        \subfloat[Energy of retrograde geodesics.]
        {
            \includegraphics[width=0.49\textwidth]{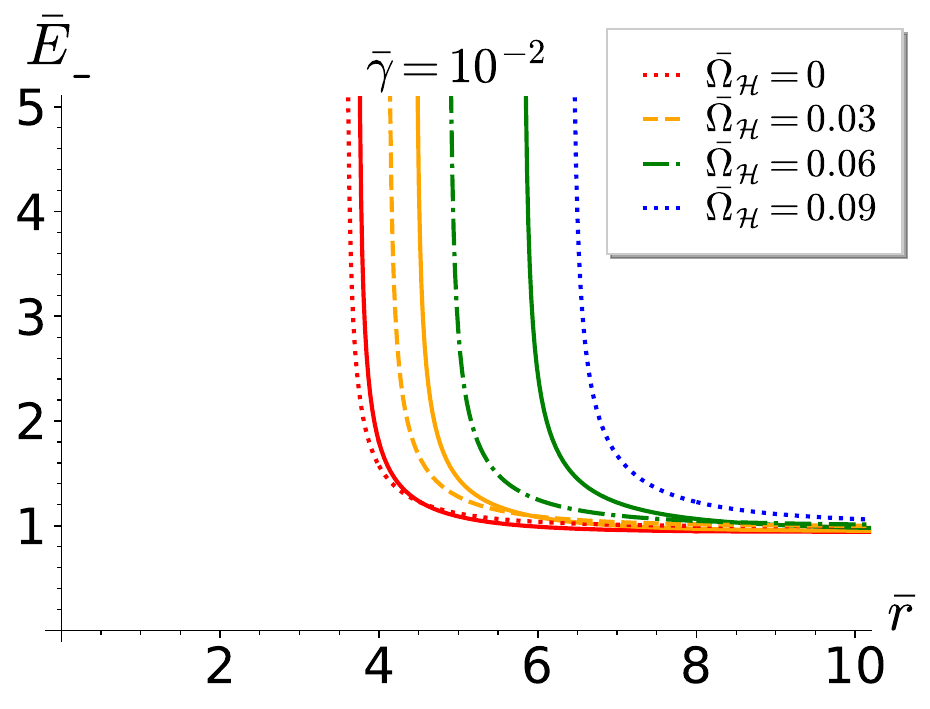}
            \label{fig_Em_rot}
        } \\
        \subfloat[Stability of prograde orbits from function~$\mathcal{V}_{+}''$.]
        {
            \includegraphics[width=0.49\textwidth]{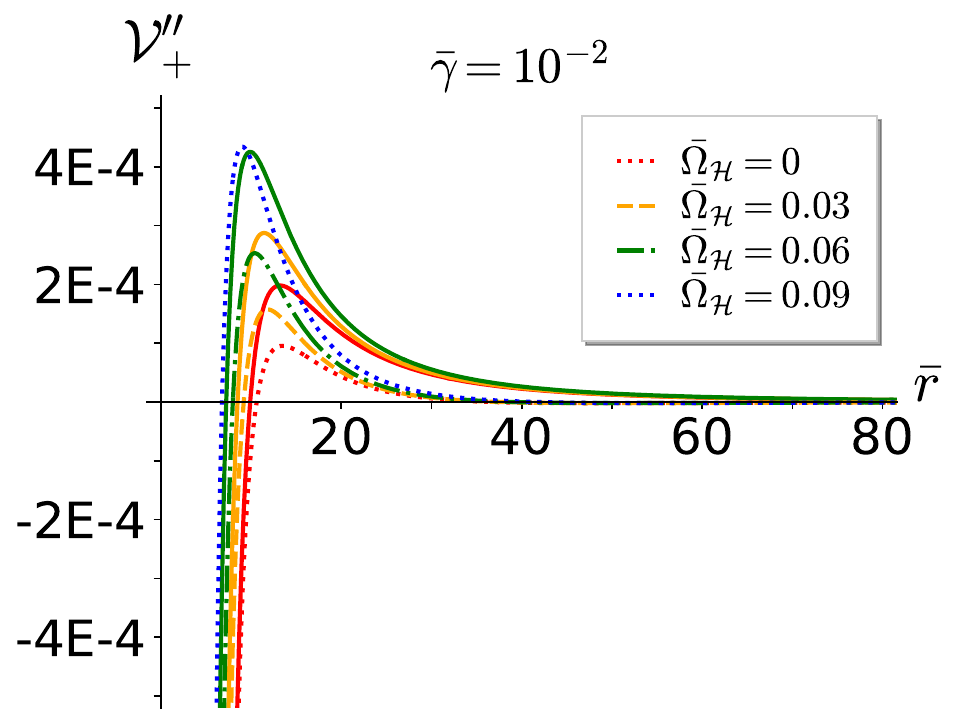}
            \label{fig_drrPotp_rot}
        }
        \subfloat[Stability of retrograde orbits from function~$\mathcal{V}_{-}''$.]
        {
            \includegraphics[width=0.49\textwidth]{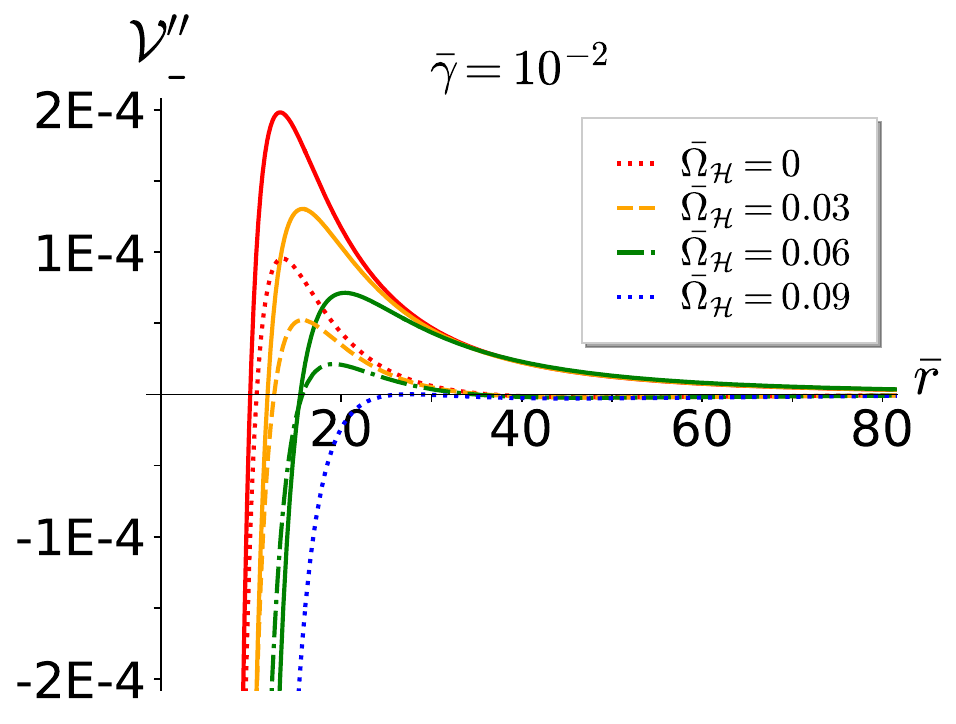}
            \label{fig_drrPotm_rot}
        }
    \end{center}
\caption{Kinematic quantities measured by the ZAMO and stability of the timelike circular geodesics for different angular velocities~$\bar{\Omega}_{\mathcal{H}} = r_{\mathcal{H}} \Omega_{\mathcal{H}}$ at fixed coupling~$\bar{\gamma}~=~10^{-2}~<~\bar{\gamma}^{c}$. For comparison, the profile in Kerr spacetime is plotted as a solid line with the same color for any fixed~$\bar{\Omega}_{\mathcal{H}}$.}
\label{fig_rot}
\end{figure}

Rotation breaks spherical symmetry so that the ``+'' and ``-'' quantities are no longer equal or opposite, as shown in figures~\ref{fig_rot} (in which solid lines correspond to the analogue quantities in Kerr spacetime).
Yet all these quantities have the same behaviour at the boundaries as in the static, spherically symmetric case.

Besides, figure~\ref{fig_V_rot} shows that~$V_{+} > 0$ and~$V_{-} < 0$ still hold everywhere.
However~$V_{+} = -V_{-}$ no longer does, so that there exist a prograde photon ring and a distinct retrograde one for each angular velocity~$\bar{\Omega}_{\mathcal{H}} = r_{\mathcal{H}} \Omega_{\mathcal{H}}$.
As in Kerr spacetime, prograde and retrograde velocities decrease as~$\bar{\Omega}_{\mathcal{H}}$ increases, so that the prograde (resp. retrograde) ring radius decreases (resp. increases).
The dependence on~$\bar{\Omega}_{\mathcal{H}}$ yet seems stronger in Kerr spacetime, meaning e.g. that the prograde ring radius decreases faster than for any non-zero~$\bar{\gamma}$.
Since the photon ring of the static, spherically symmetric Galileon spacetime is below that of Schwarzschild spacetime, the relative positions of the Kerr and Galileon prograde rings are inverted for some~$\bar{\Omega}_{\mathcal{H}}$~($\simeq 0.03$ for~$\bar{\gamma}~=~10^{-2}$).
On the contrary, the Kerr retrograde ring grows away from its Galileon counterpart.

The fact that the dependence on~$\bar{\Omega}_{\mathcal{H}}$ is qualitatively the same in Kerr and Galileon spacetimes, but stronger in the former, also applies to~$\bar{L}$ (figure~\ref{fig_L_rot}),~$\bar{E}$ (figures~\ref{fig_Ep_rot} and~\ref{fig_Em_rot}) and~$\mathcal{V}_{\pm}''$ (figures~\ref{fig_drrPotp_rot} and~\ref{fig_drrPotm_rot}).
Besides,~$\mathcal{V}_{+}''$ (resp.~$\mathcal{V}_{-}''$) globally increases (resp. decreases) as~$\bar{\Omega}_{\mathcal{H}}$ increases.
Therefore, both the prograde ISCO and retrograde OSCO (resp. prograde OSCO and retrograde ISCO) radii decrease (resp. increase) with rotation.
Interestingly, since the ISCO radius of static, spherically symmetric Galileon black holes is beyond Schwarzschild's ISCO, and Kerr's retrograde ISCO increases faster with~$\bar{\Omega}_{\mathcal{H}}$, the relative positions of the Kerr and Galileon retrograde ISCO's are inverted for some~$\bar{\Omega}_{\mathcal{H}}$~($\simeq 0.06$ for~$\bar{\gamma}~=~10^{-2}$).

\begin{figure}
    \begin{center}
        \subfloat[Stability of prograde orbits for~$\bar{\gamma}~=~1~>~\bar{\gamma}^{c}$.]
        {
            \includegraphics[width=0.49\textwidth]{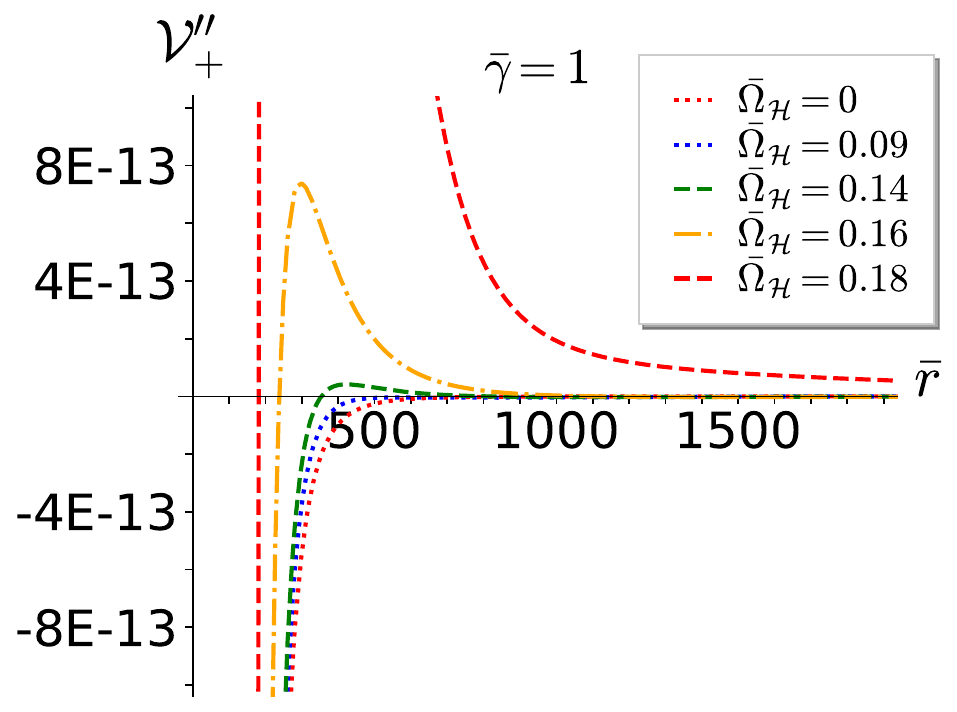}
            \label{fig_drrPotp_rot_gamma1}
        }
        \subfloat[Enlargement of figure~\ref{fig_drrPotp_rot_gamma1}.]
        {
            \includegraphics[width=0.49\textwidth]{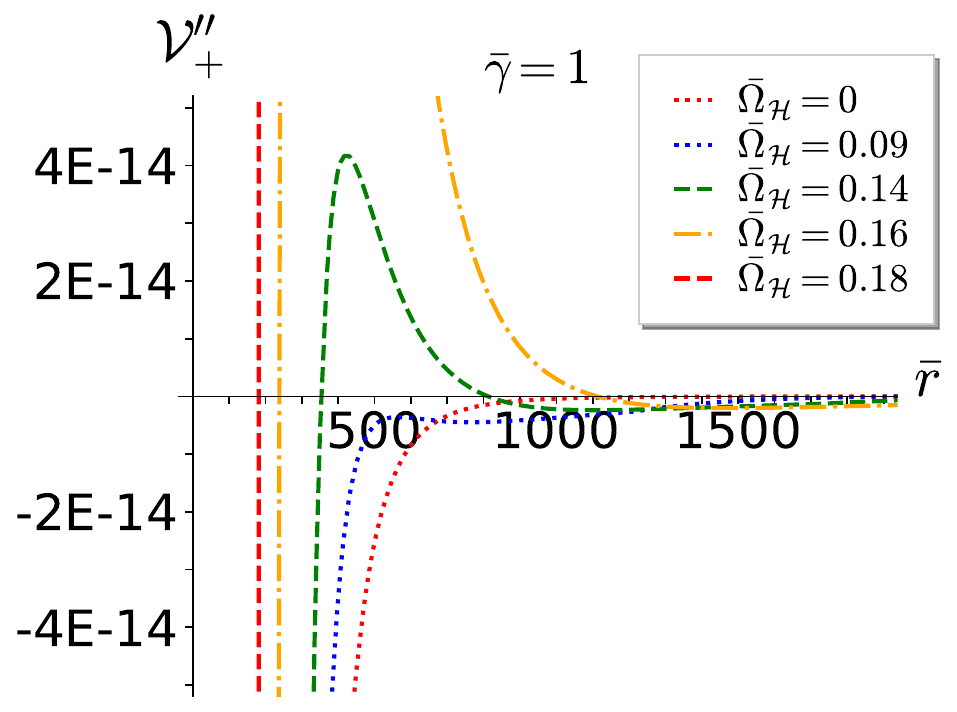}
            \label{fig_drrPotp_rot_gamma1_zoom}
        }
    \end{center}
    \caption{Rotation restores prograde stable orbits.}
\label{fig_drrPotp_rot_gamma1_and_zoom}
\end{figure}

Furthermore, sufficiently high~$\bar{\Omega}_{\mathcal{H}}$ makes it possible for~$\mathcal{V}_{+}''$ to become positive even for~$\bar{\gamma}$ greater than the critical coupling~$\bar{\gamma}^{c}~\simeq~2.2~\times~10^{-2}$; this is illustrated in figure~\ref{fig_drrPotp_rot_gamma1} for~$\bar{\gamma} = 1 > \bar{\gamma}^{c}$.
Therefore, for each coupling~$\bar{\gamma}~>~\bar{\gamma}^{c}$, there is a minimal angular velocitiy beyond which stable orbits reappear, yet only prograde ones.
On the contrary, as~$\bar{\Omega}_{\mathcal{H}}$ increases, the Galileon retrograde ISCO and OSCO eventually merge for a critical angular velocity~$\bar{\Omega}_{\mathcal{H}}^{c}$~($\simeq 0.09$ for~$\bar{\gamma}~=~10^{-2}$), beyond which no stable retrograde orbit exists anywhere.
Therefore, the fact that stars stably orbit~\emph{Sgr~A*} in both directions
\footnote{
The spin direction of~\emph{Sgr~A*} is unknown.
} 
leads to even tighter constraints than~$\bar{\gamma} \ll \bar{\gamma}^{c}$ if~\emph{Sgr~A*} is modeled as a rotating black hole.

\section{Images of cubic Galileon black holes}
\label{section_im_num}

\subsection{Principle of ray-tracing}
\label{section_ray_tracing}

In the present section, images of an accretion disk orbiting the black holes are computed numerically.
This is again motivated by the idea that the present model may capture the significant characteristics of more complete, better-behaved models.
Computations are performed by the free, open-source ray-tracing code~\emph{GYOTO}~\cite{Gyoto}, which features an efficient approach to integrate the geodesic equations from the knowledge of the 3+1 quantities decomposing a numerical metric~\cite{3p1_geod}.
In our case, the shift~$\beta$ and spatial metric~$\tensor*[^{3}]{g}{}$ corresponding to the quasi-isotropic metric~(\ref{eq_circular_spacetime}) are
\begin{eqnarray}
\beta = -\omega \partial_{\phi},               \\
\tensor*[^{3}]{g}{} = {\rm diag}(A^{2}, A^{2}r^{2}, B^{2}r^{2} \sin^{2}\theta).
\end{eqnarray}

Images are computed in the following way.
An explicit model of accretion flow is set around the black hole
\footnote{
Rough estimates, confirmed by simple exact models of accretion disks,
show that the gravitational influence of an accretion disk is usually completely negligible with respect to the black hole.
Thus, the vacuum black hole metrics are still valid in presence of an accretion disk.
See section~6.5 of~\cite{Bambi_BH_test_strong_grav_book} for quantitative arguments.
}
. 
A telescope set in the numerical metric mimicks the observing wavelength~($1.3 \, \text{mm}$), the distance~($16.9 \, \text{Mpc}$), field of view~($120 \, \mu\text{as}$) and orientation of the Event Horizon Telescope with respect to~\emph{M87*}~(the black hole being set at the origin and the disk lying in the equatorial plane~$\theta = 90^{\circ}$, the colatitude of the Earth is~$\theta = 160^{\circ}$, while the vertical axis of the screen of the EHT is rotated by $110^{\circ}$ clockwise from the projection on the screen of the spin axis of the disk).
Each pixel of its focal screen corresponds to a spatial direction, which uniquely defines the initial tangent vector of a null affinely parametrized geodesic.
The latter is integrated backwards in time until a stopping condition is met, e.g. the photon gets too close to the event horizon, or definitely leaves the strong field region.
Otherwise, every time the geodesic crosses the accretion disk, the radiative transfer equations ruling the specific intensity are integrated along the segment lying within the disk.
The cumulated specific intensity is eventually plotted on the initial pixel.

Yet, determining the nature and properties of a compact object based on the image of its accretion flow is a very degenerate inverse problem~\cite{Bambi19_shadow_bh_mimicker, Shaikh19_image_naked_singu, Glampedakis21_shadow_test_Kerr}.
This is for instance evidenced in reference~\cite{image_boson} in which the same model of accretion disk is set around a boson star and a black hole: the deviations between the resulting images are very subtle although the natures of the accreting objects are very different.
Furthermore, the resolution of present and future instruments like the EHT is limited, making it even harder to distinguish subtle features
\footnote{
Regarding the particular problem of distinguishing boson stars from black holes, see however~\cite{Olivares20_accret_boson} for possibly detectable deviations arising from dynamical effects revealed by 3D general relativistic magnetohydrodynamics simulations.
}
.

Then, the purpose of numerical images is not systematically to check whether the image constructed by the EHT~\cite{EHT_Shadow_M87} can be reproduced for different accreting compact objects.
This indeed requires costly general relativistic magnetohydrodynamics (GRMHD) simulations, together with a model of the EHT itself.
Instead, strong efforts are made to propose fairly simple and yet realistic models of accretion disks~\cite{Vincent_ion_tor, Abramowicz_review_accretion, Vincent_magn_torus, Vincent_model_M87_2020}.
In particular, such models are assumed to be good approximations of stable steady solutions of the GRMHD equations.
Comparing the resulting images for different compact objects provides a more efficient and still relevant method to evaluate how degenerate the problem is.
The hope is that such an approach should help isolating the causes of differences between images, e.g. being able to guess the nature and amplitudes of the modifications that result from changing the accretion model or the theory used to describe the whole system.

\setcounter{footnote}{0} 

As a result, a simple model of accretion disk, recently introduced in~\cite{Vincent_model_M87_2020}, is used in the sections below.
Like~\emph{Sgr~A*}, supermassive black hole~\emph{M87*} features a very low-luminosity accretion flow, revealing an inefficient radiative cooling and hence a high temperature.
It is consistently modeled as a low accretion rate, geometrically thick, optically thin disk
\footnote{
An accretion disk is geometrically (resp. optically) thin when the opening angle (resp. optical depth) is smaller than~$1$. It is geometrically (resp. optically) thick otherwise.
}
~\cite{Narayan_review}.
Besides these properties, only the thermal synchrotron emission is computed, following a method exposed in~\cite{Pandya_2016}.
In the end, the complete model is described by very few input parameters: the opening angle and inner edge of the disk (which is set at the ISCO in our case), the magnetization parameter (which determines the ambient magnetic field strength), the electron density and temperature at the inner edge (which determine the density and temperature profiles).

\subsection{Static and spherically symmetric case}
\label{section_im_stat}

\begin{figure}
    \begin{center}
        \subfloat[$(\bar{\gamma},\bar{\Omega}_{\mathcal{H}}) = (0,0)$ (Schwarzschild)]
        {
            \includegraphics[width=0.5\textwidth]{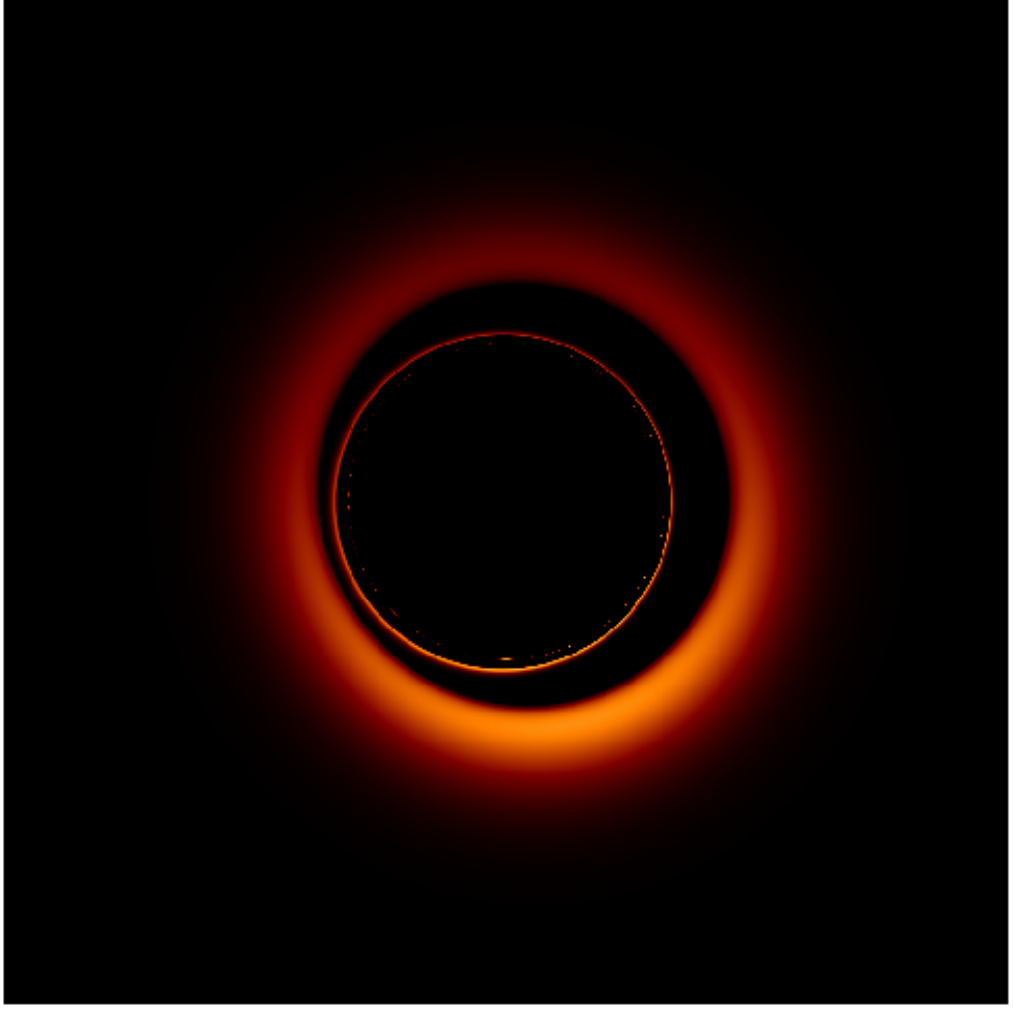}
            \label{fig_merged_stat_Schw_DT33.8_FoV120_RES1000_scale0to7e-7_noScaleBar}
        }
        \subfloat[$(\bar{\gamma},\bar{\Omega}_{\mathcal{H}}) = (10^{-2},0)$ (static Galileon)]
        {
            \includegraphics[width=0.5\textwidth]{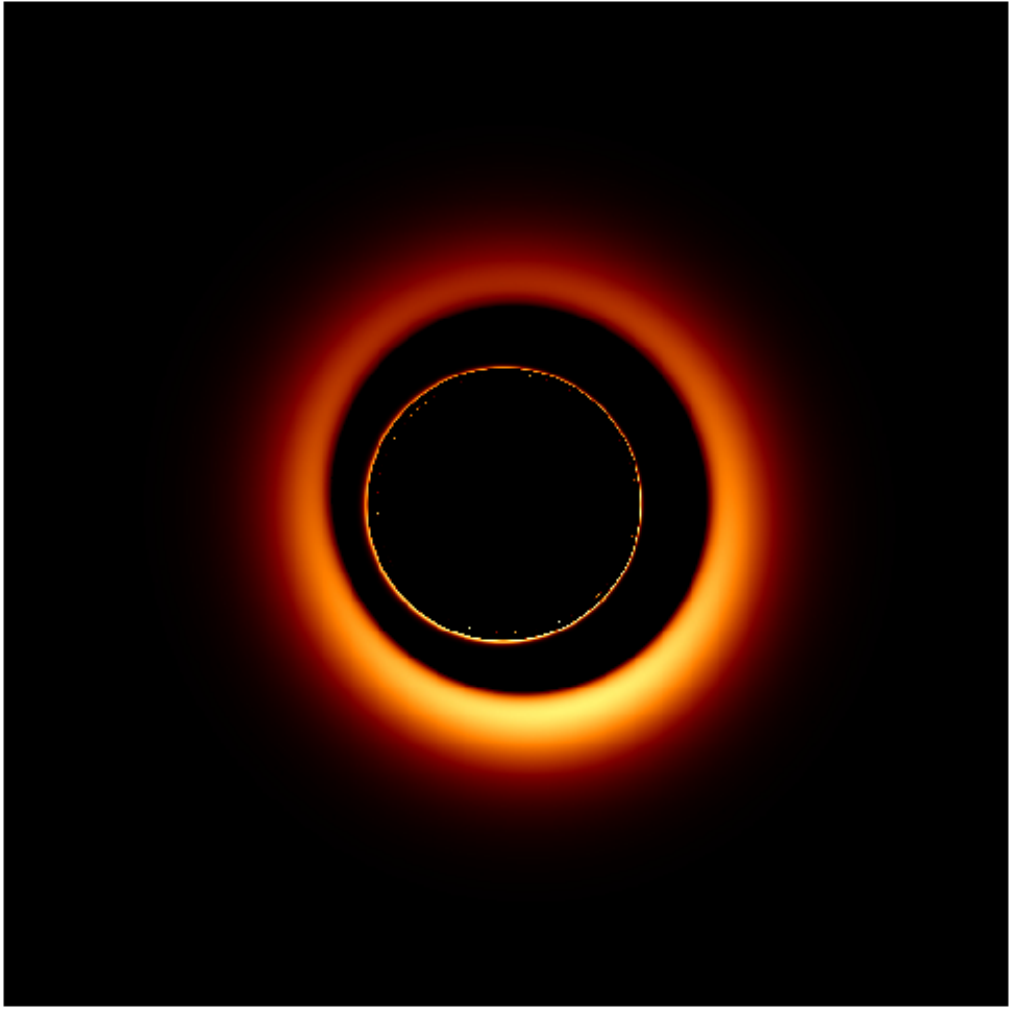}
            \label{fig_merged_stat_Gal_a1a2.01_DT33.8_FoV120_RES1000_scale0to7e-7_noScaleBar}
        }   \\
        \subfloat[$(\bar{\gamma},\bar{\Omega}_{\mathcal{H}}) = (0,0.07)$ (Kerr)]
        {
            \includegraphics[clip, trim=0.4cm 0cm 0.4cm 0cm, width=0.5\textwidth]{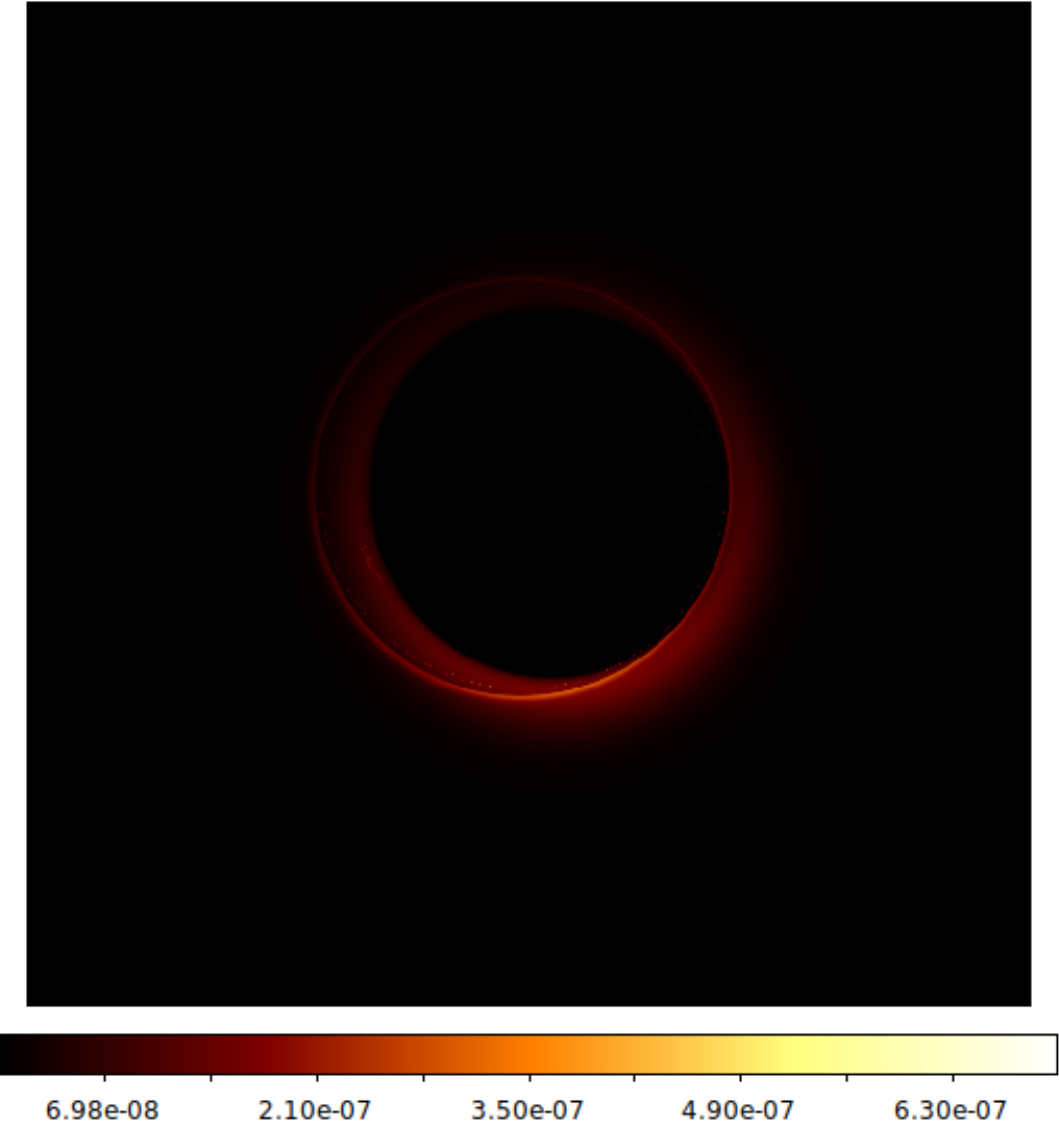}
            \label{fig_merged_rot_Kerr_adOm.07_DT33.8_FoV120_RES1000_scale0to7e-7}
        }
        \subfloat[$(\bar{\gamma},\bar{\Omega}_{\mathcal{H}}) = (10^{-2},0.07)$ (rotating Galileon)]
        {
            \includegraphics[clip, trim=0.4cm 0cm 0.4cm 0cm, width=0.5\textwidth]{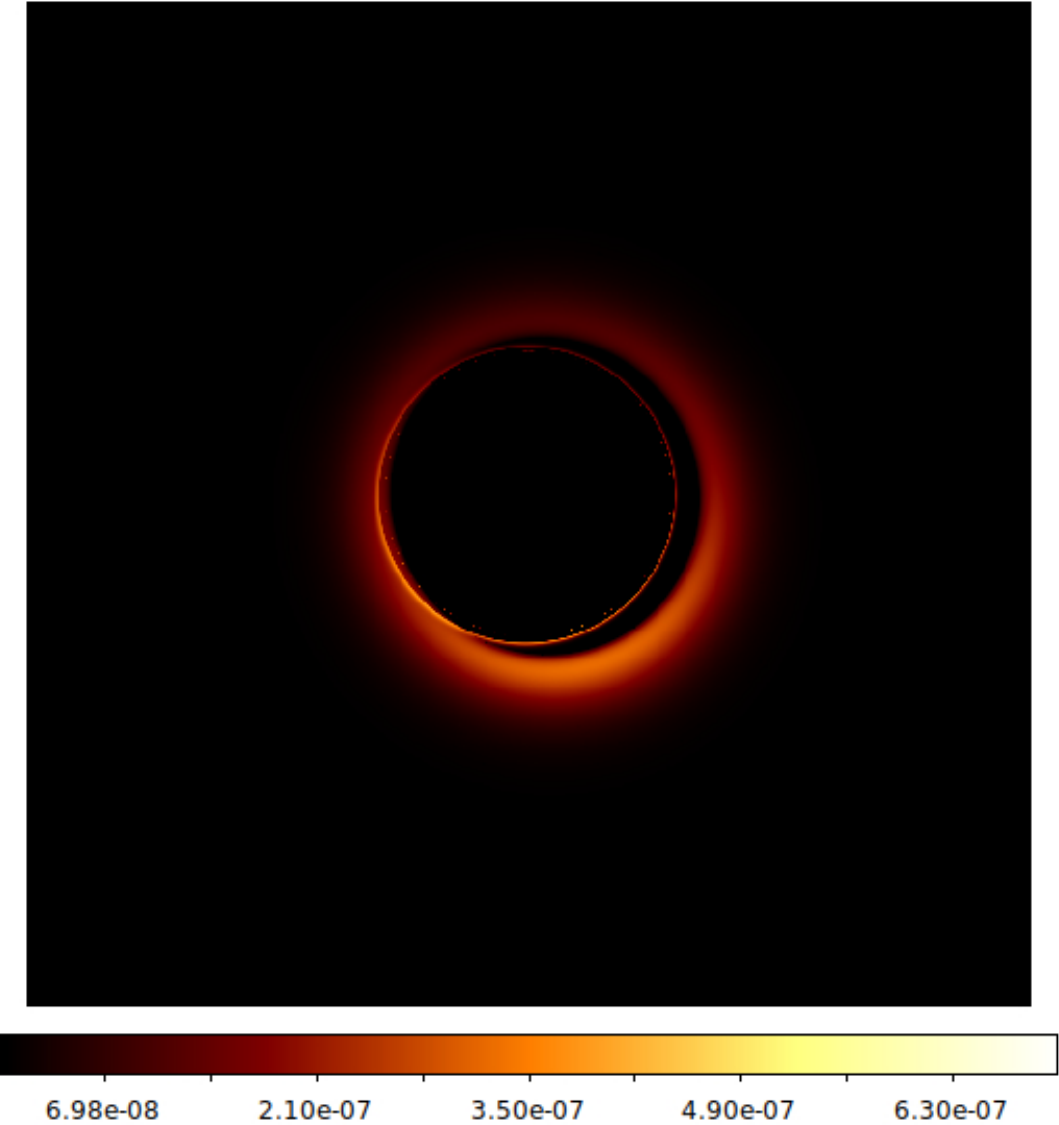}
            \label{fig_merged_rot_Gal_a1a2.01_adOm.07_DT33.8_FoV120_RES1000_scale0to7e-7}
        }
    \end{center}
\caption{Images at~$1.3 \, \text{mm}$ produced by a thick accretion disk orbiting black holes for various couplings~$\bar{\gamma}$ and angular velocity~$\bar{\Omega}_{\mathcal{H}}$ (fixed horizon radius~$r_{\mathcal{H}} = M_{\emph{M87*}}/2$). The field of view equals~$120 \, \mu\text{as}$. The linear scale of the specific intensity~$I_{\nu}$ is provided in Systeme International (SI) units ($I_{\nu} = 4 \times 10^{7} \, \text{SI}$ corresponds to a brightness temperature~$\sim25 \times 10^{9} \, \text{K}$).
}
\label{fig_image_BH}
\end{figure}

To connect with section~\ref{section_geod_num}, the black holes on figure~\ref{fig_image_BH} all have the same radius, namely the Schwarzschild radius of~\emph{M87*} (in quasi-isotropic coordinates, i.e.~$r_{\mathcal{H}} = M_{\emph{M87*}}/2$ where~$M_{\emph{M87*}} \simeq 6.5 \times 10^{9} M_{\odot}$).
It is actually more natural to manipulate the horizon radius than any notion of mass: because of the non-Schwarzschild asymptotics of the Galileon black holes recalled in section~\ref{section_geod_stat}, there is no relevant mass parameter that can be extracted from star trajectories (through Kepler's law) or surface integrals (such as the Komar or Arnowitt-Deser-Misner masses) in the weak field region, since its value could not be compared in any meaningful way to that of a Schwarzschild black hole (recall for instance that these Galileon black holes feature a vanishing Komar mass at spatial infinity, and yet admit stable orbits).
It is thus equally legitimate to make comparisons based on parameters specific to the strong field region such as the horizon, lightring and ISCO radii, and the characteristics of the images described below.

In the non-rotating case, images~\ref{fig_merged_stat_Schw_DT33.8_FoV120_RES1000_scale0to7e-7_noScaleBar} and~\ref{fig_merged_stat_Gal_a1a2.01_DT33.8_FoV120_RES1000_scale0to7e-7_noScaleBar} compare the Schwarzschild limit~$\bar{\gamma} = 0$ to a static and spherically symmetric Galileon black hole below the critical coupling.
In both cases, the ISCO radius is read from figure~\ref{fig_drrPot_stat}.
The asymmetries within each image come from the configuration of the EHT with respect to the accretion disk.
More precisely, the $110^{\circ}$ clockwise rotation of the vertical axis of the screen from the projected spin axis of the disk explains the position on the images of the brighter spot that results from the relativistic beaming and blueshift affecting the part of the disk rotating towards the screen.
Besides, the inside luminous ring corresponds to the secondary and higher order images, which aymptotically accumulate in the direction of the lightring.
Rather than being centered within the primary image of the disk, this ring appears shifted towards the top of the spin axis because of the inclination angle $\theta = 160^\circ$ of the EHT observer
(the disk is almost seen from below).

This ring gets smaller with respect to the primary image as~$\bar{\gamma}$ increases because the ISCO radius increases (figure \ref{fig_drrPot_stat}) while the lightring decreases (figure \ref{fig_V_stat}).
Explicitly, in Schwarzschild case, the internal diameter of the primary image is~$D \simeq 50 \, \mu\text{as}$ while the diameter of the secondary ring is~$d \simeq 42 \, \mu\text{as}$, which is compatible with the EHT image~\cite{EHT_Shadow_M87}.
In the Galileon case,~$D \simeq 46 \, \mu\text{as}$ while~$d \simeq 34 \, \mu\text{as}$.
Besides, the image is globally brighter as~$\bar{\gamma}$ increases.
This is consistent with the fact that asymptotic convergence to Minkowski is faster as~$\bar{\gamma}$ increases (as recalled in section~\ref{section_geod_stat}).
In other terms, the strong field region shrinks as~$\bar{\gamma}$ increases, so that most light rays undergo a weaker gravitational redshift.

Therefore, if one considered a Galileon black hole with e.g. a~$1.12$ times greater radius, so as to fit the internal and secondary diameters close enough to the Schwarzschild values (the deviation on their ratio being possibly undetectable by the resolution of the EHT
\footnote{
Actually, in order to reproduce even more closely the diameters, and hence their ratio, one could fine tune the dimension of the internal diameter of the primary image by considering a more realistic scenario involving accreting matter on non-circular trajectories below the ISCO. In this sense, the secondary ring is more reliable as an observable of the gravitational field as it weakly depends on the boundaries and physical properties of the accretion process.
}
), the corresponding image would get even brighter than figure~\ref{fig_merged_stat_Gal_a1a2.01_DT33.8_FoV120_RES1000_scale0to7e-7_noScaleBar}.
One could accordingly reduce the density parameter to avoid tension with the observed luminosity.
Yet the fact would remain that the ISCO radius would be even greater than it already is with respect to its Schwarzschild analogue when the horizon radii are the same (figure~\ref{fig_drrPot_stat}).
It is hardly conceivable that the ISCO radius of~\emph{M87*} could be estimated by other means than EHT-type observations, so that no incompatiblity related to this parameter could be exhibited.
On the other hand, if images of~\emph{Sgr A*} were obtained, tensions about the ISCO location should arise based on the high precision astrometric observations made by the instrument GRAVITY which detected ``flares''~\cite{GRAVITY_motion_ISCO} close to Schwarzschild ISCO, knowing that such bright spots are expected to materialize near the inner edge of the accretion disk of~\emph{Sgr A*}.

\subsection{Rotating case}
\label{section_im_rot}

\begin{figure}
    \begin{center}
    \includegraphics[clip, trim=0.7cm 0cm 0.7cm 0.4cm, width=0.5\textwidth]{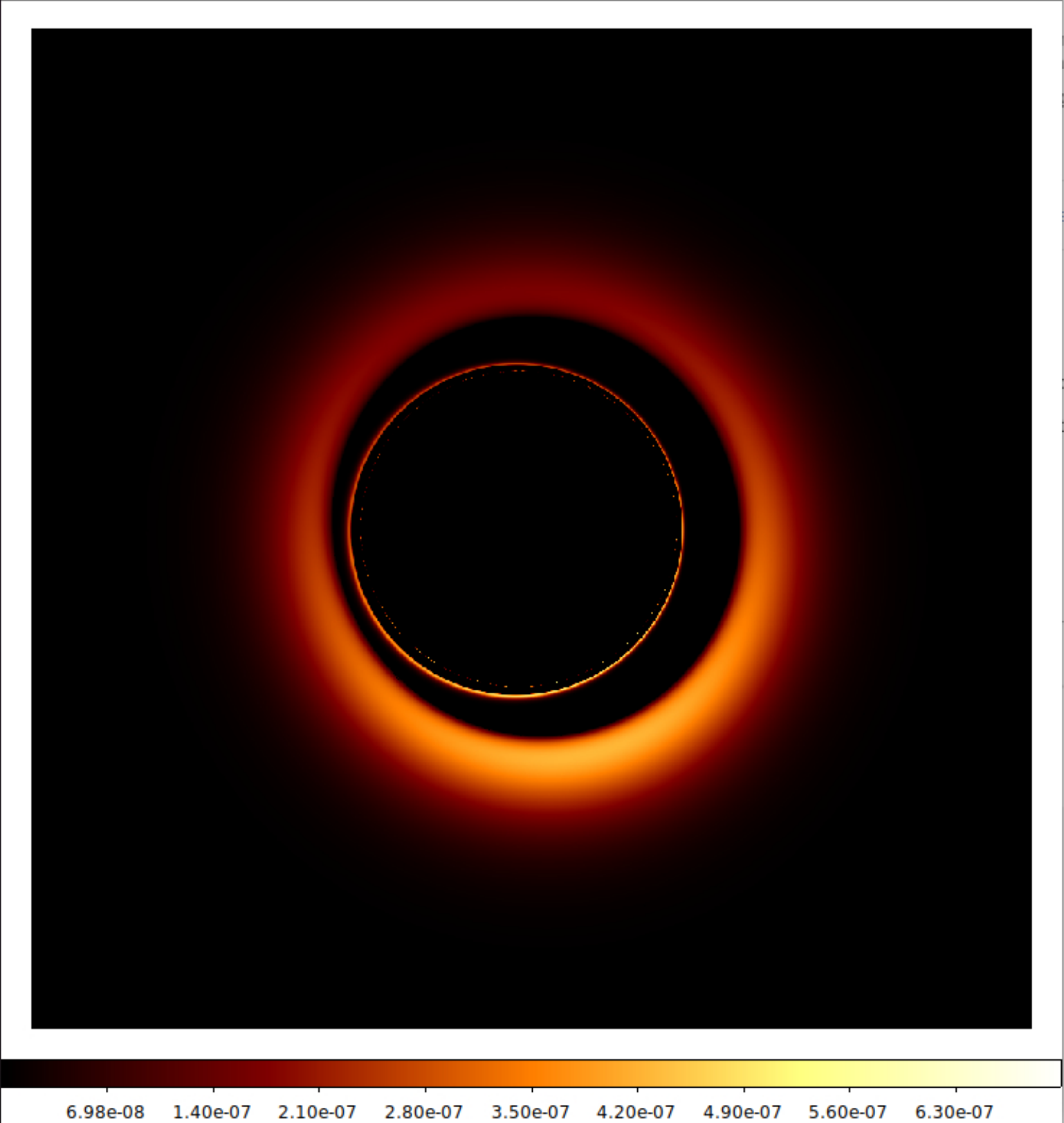}
    \end{center}
    \caption{Case~$(\bar{\gamma},\bar{\Omega}_{\mathcal{H}}) = (10^{-2},0.03)$ such that~$r_{\mathcal{H}} = 1.2 M_{\emph{M87*}}/2$. To compare with figure~\ref{fig_merged_stat_Schw_DT33.8_FoV120_RES1000_scale0to7e-7_noScaleBar}.}
\label{fig_merged_rot_Gal_a1a2.01_adOm.03_DT27.98_FoV120_RES1000_scale0to7e-7}
\end{figure}

Images~\ref{fig_merged_rot_Kerr_adOm.07_DT33.8_FoV120_RES1000_scale0to7e-7} and~\ref{fig_merged_rot_Gal_a1a2.01_adOm.07_DT33.8_FoV120_RES1000_scale0to7e-7} compare a Kerr black hole to a rotating Galileon black hole sharing the same angular velocity.
In both cases, the inner edge of the disk is set at the prograde ISCO, which is read from figure~\ref{fig_drrPotp_rot}.
Besides, the inside ring intersects the primary image, meaning that the ISCO is close enough to the light ring to allow photons to cross the disk at least a second time.
This is consistent with the fact that the prograde ISCO decreases (figure \ref{fig_drrPotp_rot}) faster than the prograde lightring (figure \ref{fig_V_rot}) as~$\bar{\Omega}_{\mathcal{H}}$ increases.
As in the static case, the image is globally brighter as~$\bar{\gamma}$ increases.
But at a fixed~$\bar{\gamma}$, the images get darker as~$\bar{\Omega}_{\mathcal{H}}$ increases because the strong field region expectedly expands with rotation.

As illustrated by figure~\ref{fig_merged_rot_Gal_a1a2.01_adOm.03_DT27.98_FoV120_RES1000_scale0to7e-7}, such tendencies leave room for a black hole with moderately low rotation, and radius greater than~$M_{\emph{M87*}}/2$, to closely mimick Schwarzschild image~\ref{fig_merged_stat_Schw_DT33.8_FoV120_RES1000_scale0to7e-7_noScaleBar} in terms of dimensions and brightness.
Yet regarding the ISCO radius, the fit is expectedly not so good with respect to Schwarzschild ($9\%$ and~$50\%$ deviation on the prograde and retrograde ISCO respectively).
Of course, clearing away degenerate tendencies and exhibiting significant incompatibilities would only be possible if images and more precise spin measurements of~\emph{Sgr A*} were available.

\section{Conclusion}

Investigating the geodesics of asymptotically flat cubic Galileon black holes exhibits non-viable characteristics when the coupling~$\bar{\gamma}$ is in a range leading to non-negligible metric deviations from GR.
As such, the model would be constrainted to a negligible coupling~$\bar{\gamma}$ for the OSCO to lie far enough to be compatible with distant stars observed around~\emph{Sgr~A*}.
Yet it is thought that constraints only based on the OSCO might be dismissed by restoring terms in Lagrangian~(\ref{eq_action}), or invoking other appropriate fields, which would preserve the tendencies and strong-field characteristics of the asymptotically flat model.
This is why images of accretion disks have also been computed, as these allow to probe the close environment of a supermassive black hole.
The dependence of this observable on coupling and rotation may lead to deviations from GR in terms of global brightness and relative dimensions of the luminous structures.
Yet these may often be compensated by adjusting density and the horizon radius.

Proper constraints will hopefully arise from future images and precise measurements of the spin and ISCO location of~\emph{Sgr A*}.
The latter being one of the main targets of the EHT, knowledge about this system will keep increasing in the coming years.
Besides probing the inner accretion flow~\cite{Pu18EHT_Prob_accret_geom_SgrA, Bower18EHT_Prob_accret_SgrA_Bondi}, the observations from the EHT will contribute to evaluating its spin (while former estimations are based e.g. on quasi-peridodic oscillations of the radio emission~\cite{Kato10_SpinSupermassiveBH_SgrA}).
Indeed, EHT data was already used to establish bounds on the spin of~\emph{M87*}~\cite{Nemmen19_spinM87, Tamburini19_Meaure_spinM87_twisted_light}, and this will improve with further observations from the EHT~\cite{EHT21_Constraints_BHcharges_M87}.
Information on the spin may also be extracted from instruments of the Very Large Telescope such as the interferometer GRAVITY which can monitor faint stars very close to~\emph{Sgr A*}~\cite{Fragione20_UpperLimSpinSgrA}.
From a broader point of view, GRAVITY also contributes to better understanding the astrophysics of massive black holes~\cite{GRAVITY20_hot_dust_AGNs}, and such general knowledge will also improve in future decades thanks to space-based gravitational-wave observatory LISA~\cite{Barack04_LISA_capture_sources, Babak17_LISA_EMRI}, which will extract precious and precise information from stellar-mass compact objects spiralling around massive black holes.
On an even longer term, reference~\cite{Gralla20_spaceVLBI} proposes a space-based very long baseline interferometry experiment which would characterize the ring-shaped structures in images of black hole accretion flows with extremely high precision, measuring the diameter size to $0.04\%$ accuracy, while the EHT is limited to approximately $10\%$ accuracy.
This highlights again the fact that constraining theories of gravity with massive black holes is a long-term enterprise, as determining the parameters of black holes and their accretion flow is a degenerate problem conditioned by instrumental accuracy and modeling limitations~\cite{Bambi20_review_astroBH, Zhu18_TestGR_SgrA_Limits_Scatter, Issaoun21_NonGaussian_SgrA}.
It is in particular delicate to predict when and which opportune combination of observations will provide inflexible constraints or could at least strongly disfavour a model.

The models considered in the present paper are not free of simplifications either, both in the metric and the model of disk, which could be addressed in later work.
As explained in the introduction, the circular metric~(\ref{eq_circular_spacetime}) is only exact in the static case.
Reproducing the rotating solutions in a noncircular framework would allow to reach accurate rapidly rotating solutions (possibly up to an extremal case).
Such general framework would also be used to construct the asymptotically de Sitter configurations corresponding to non-zero~$\eta$ and~$\Lambda$, and perform similar study and comparisons.
Furthermore, in order to make precise quantitative comparisons between actual images and numerical predictions in a consistent way, more realistic models of disk such as ion tori could be considered.
The latter are also geometrically thick and optically thin structures, yet featuring more complex density and temperature profiles derived from first principles, as well as isotropic~\cite{Vincent_ion_tor} or toroidal~\cite{Vincent_magn_torus} magnetic fields, hence allowing to explore other parts of the parameter space describing accretion disks.

\section*{Acknowledgements}

KV and EG acknowledge support from the CNRS program 80PRIME-TNENGRAV.

\appendix

\section{Equatorial geodesics in quasi-isotropic coordinates}
\label{appdx_geod_QI}

The present appendix recalls useful results on particles freely moving in the equatorial plane of a circular spacetime, described in terms of the quasi-isotropic coordinates~(\ref{eq_circular_spacetime}) (see~\cite{Bardeen_stability, Bardeen_rot_bh, note_geods} and sections 4.6 and 4.7 of~\cite{Eric_intro_relat_stars} for closely related discussions).
Based on the unique parametrization
\begin{eqnarray}
\label{eq_curve}
\mathcal{C}:\ \lambda \mapsto \left( x^{\mu}(\lambda) \right) = \left( t(\lambda),\ r(\lambda),\ \pi/2,\ \phi(\lambda) \right)
\end{eqnarray}
such that the 4-momentum of the particle is~$p^{\mu} = \dot{x}^{\mu}$ (where a dot denotes differentiation with respect to the parameter~$\lambda$), the geodesic equation
\begin{eqnarray}
\label{eq_geodesic}
\nabla_{p}p = 0
\end{eqnarray}
implies that
\begin{eqnarray}
\label{eq_E_conservation}
E = - p_{t} \ \ \text{ is conserved along } \mathcal{C}, \\
\label{eq_L_conservation}
L = p_{\phi} \ \ \text{ is conserved along } \mathcal{C}, \\
\label{eq_equatorial_condition}
\theta = \pi/2\ \ \text{ is conserved along } \mathcal{C}, \\
\label{eq_radial}
\frac{\dot{r}^{2}}{2} + \mathcal{V}(r,m,E,L) = 0,
\end{eqnarray}
where~$m = \sqrt{- p^{2}}$ defines the mass of the particle, and the effective potential~$\mathcal{V}$ is defined as
\begin{eqnarray}
\label{eq_potential}
\mathcal{V}(r,m,E,L) = \frac{1}{2A^{2}}\left[ m^{2} - \left( \frac{E - \omega L}{N} \right)^{2} + \left( \frac{L}{B r} \right)^{2} \right].
\end{eqnarray}

Equation~(\ref{eq_radial}) is merely an explicit version of the mass conservation equation ($m$ is conserved along~$\mathcal{C}$) taking advantage of the three other conservation equations~(\ref{eq_E_conservation}),~(\ref{eq_L_conservation}) and~(\ref{eq_equatorial_condition}).
Thus, these four conservation equations are four necessary conditions for a curve~$\mathcal{C}$ to describe an equatorial trajectory of a free particle.
For non-circular orbits (i.e.~$\dot{r} \neq 0$ almost everywhere), they are sufficient and have a unique solution provided that the initial sign of~$\dot{r}$ is fixed and the initial coordinates~$(t_{0},r_{0},\pi/2,\phi_{0})$, the Killing energy~$E$, the Killing angular momentum~$L$ and the mass~$m$ satisfy~$\mathcal{V}(r_{0},m,E,L) < 0$ and~$E - \omega(r_{0}) L > 0$ (to guarantee that the trajectory is initially causal, i.e. future-oriented).
For a circular geodesic at radial coordinate~$r_{0}$, one has~$\dot{r} = 0$ so that~$\mathcal{V}(r_{0},m,E,L) = 0$.
Yet, as detailed in~\cite{note_geods}, the additional condition
\begin{eqnarray}
\label{eq_circular_geod_condition}
\mathcal{V}'(r_{0},m,E,L) = 0
\end{eqnarray}
is required besides equations~(\ref{eq_E_conservation}),~(\ref{eq_L_conservation}) and~(\ref{eq_equatorial_condition}) to realize a geodesic instead of an arbitrary (possibly accelerated) circular orbit.
Rather than~$E$ and~$L$, this constraint is usually formulated in terms of the spatial velocity~$V$ measured by the zero angular momentum observer (ZAMO)\footnote{The ZAMO are characterized by a 4-velocity collinear to~$\nabla t$.} as
\begin{eqnarray}
\label{eq_V_circu}
\left( \frac{B'}{B} + \frac{1}{r} \right) V^{2} - \frac{B r \omega'}{N} V - \frac{N'}{N} = 0.
\end{eqnarray}

Roots exist if and only if the discriminant
\begin{eqnarray}
\label{eq_D}
D = \left( \frac{B r \omega'}{N} \right)^{2} + \frac{4N'}{N} \left( \frac{B'}{B} + \frac{1}{r} \right)
\end{eqnarray}
is non-negative, in which case one has
\begin{eqnarray}
\label{eq_V_pm}
V_{\pm}(r) = \frac{\frac{B r \omega'}{N} \pm \sqrt{D}}{2\left( \frac{B'}{B} + \frac{1}{r} \right)}.
\end{eqnarray}

A timelike circular geodesic (resp. photon ring) exists at~$r$ when~$V_{\pm}$ are defined at~$r$ and at least one of them belongs to~$(-1,1)$ (resp.~$\{-1,1\}$) since the ZAMO necessarily measures a subluminal (resp. luminal) velocity.
The corresponding Killing energy and angular momentum of the geodesic are
\begin{eqnarray}
\label{eq_L_circu_V}
L_{\pm} = \mathcal{E} B r V_{\pm}, \\
\label{eq_E_circu_V}
E_{\pm} = \mathcal{E} ( N + B r \omega V_{\pm} ),
\end{eqnarray}
where~$\mathcal{E}$ is the energy measured by the ZAMO: for a massive particle,~$\mathcal{E} = \Gamma m$ where~$\Gamma$ is the Lorentz factor of the particle with respect to the ZAMO, while for a massless particle,~$\mathcal{E} = h \nu$ where~$\nu$ is the frequency measured by the ZAMO.

Finally, the radial equation~(\ref{eq_radial}) expectedly provides a stability criteria for circular geodesics based on convexity: for any perturbation to be bounded in some neighbourhood of a geodesic at~$r$,~$\mathcal{V}''(r,m,E,L)$ must be positive.
Since the values of~$E$ and~$L$ for a circular geodesic at~$r$ are necessarily given by relations~(\ref{eq_L_circu_V}) and~(\ref{eq_E_circu_V}), one only has to study the sign of the two functions
\begin{eqnarray}
\label{eq_sign_stab}
\mathcal{V}_{\pm}'': r \mapsto \mathcal{V}''\left(r,m,E_{\pm}(r),L_{\pm}(r)\right)
\end{eqnarray}
on the set on which the discriminant~$D$ is non-negative.
Actually, the expressions~$\mathcal{V}_{\pm}''(r)$ are homogeneous with respect to~$\mathcal{E}$, so that their sign do not depend on~$\Gamma_{\pm} m$ in the massive case nor on~$h \nu$ in the massless case.
Therefore, the stability of the causal circular geodesics only depends on the sign of the two functions~(\ref{eq_sign_stab}) and corresponds to massive particles where~$V_{\pm}(r)~\in~(-1,1)$ and massless ones where~$V_{\pm}(r)~=~\pm 1$, regardless of whether the expressions used for~$E_{\pm}$ and~$L_{\pm}$ apply to a massive or a massless particle.
These are therefore the two functions that are plotted in section~\ref{section_geod_num} to study the stability of circular geodesics in the cubic Galileon spacetimes.

\section*{References}

\bibliographystyle{iopart-num}
\bibliography{biblio}

\end{document}